\documentclass[onecolumn]{aastex62}

\usepackage{amssymb}
\usepackage{color}
\usepackage{amsmath}
\usepackage{tablefootnote}
\usepackage{comment}
\usepackage{lipsum}
\usepackage{hyperref}
\usepackage{soul}
\usepackage{graphicx}
\usepackage{natbib}
\usepackage{comment}
\usepackage{multirow}
\begin{document}

\newcommand{\Var}{\operatorname{Var}}
\newcommand{\Cov}{\operatorname{Cov}}
\newcommand{\Avg}{\operatorname{Avg}}
\newcommand{\E}{\operatorname{E}}

\title{Improved Measurements of Galaxy Star Formation Stochasticity from the Intrinsic Scatter of Burst Indicators}
\author[0000-0002-7767-5044]{Adam Broussard}
\affiliation{Department of Physics and Astronomy, Rutgers, The State University of New Jersey, 136 Frelinghuysen Rd, Piscataway, NJ 08854, USA}
\email{adamcbroussard@physics.rutgers.edu}

\author[0000-0003-1530-8713]{Eric Gawiser}
\affiliation{Department of Physics and Astronomy, Rutgers, The State University of New Jersey, 136 Frelinghuysen Rd, Piscataway, NJ 08854, USA}
\email{gawiser@physics.rutgers.edu}

\author[0000-0001-9298-3523]{Kartheik Iyer}
\affiliation{The Dunlap Institute for Astronomy and Astrophysics, 50 St. George Street, Toronto, Ontario, Canada M5S 3H4}
\email{kartheik.iyer@dunlap.utoronto.ca}

\date{June 2022, Submitted to ApJ}

\begin{abstract}

Measurements of short-timescale star formation variations (i.e., ``burstiness'') are integral to our understanding of star formation feedback mechanisms and the assembly of stellar populations in galaxies.  We expand upon the work of \cite{Broussard2019} by introducing a new analysis of galaxy star formation burstiness that accounts for variations in the $Q_{sg}=\mathrm{E(B-V)_{stars}}/\mathrm{E(B-V)_{gas}}$ distribution, a major confounding factor.  We use Balmer decrements from the MOSFIRE Deep Evolution Field (MOSDEF) survey to measure $Q_{sg}$, which we use to construct mock catalogs from the Santa Cruz Semi-Analytic Models and {\sc Mufasa} cosmological hydrodynamical simulation based on 3D-HST, Fiber Multi-Object Spectrograph (FMOS)-COSMOS, and MOSDEF galaxies with H$\alpha$ detections.  The results of the mock catalogs are compared against observations using the burst indicator $\eta = \log_{10}(\mathrm{SFR_{H\alpha}/SFR_{NUV}})$, with the standard deviation of the $\eta$ distribution indicating burstiness.  We find decent agreement between mock and observed $\eta$ distribution shapes; however, the FMOS-COSMOS and MOSDEF mocks show a systematically low median and scatter in $\eta$  in comparison to the observations.  This work also presents the novel approach of analytically deriving the relationship between the intrinsic scatter in $\eta$, scatter added by measurement uncertainties, and observed scatter, resulting in an intrinsic burstiness measurement of $0.06-0.16$ dex.

\end{abstract}


\section{Introduction}

Star formation is a key component of the formation and evolution of galaxies.  Galaxy star formation histories (SFHs) bear imprints of the various physical processes that regulate their star formation across a broad range of timescales.  Some examples of these processes and their relevant timescales include supermassive black hole feedback \citep[][; $\gtrsim 1$ Gyr]{Angles2017}, inflows and outflows of gas \citep[][; $100-500$ Myr]{Pillepich2017,Weinberger2017}, galaxy mergers \citep[][; $10-100$ Myr]{Dave2017}, and supernova feedback \citep[][; $1-20$ Myr]{Kortgen2016}.  An understanding of the variation in galaxy SFHs is central to understanding the role these processes play in galaxy evolution.

Analyses of star formation ``burstiness'' focus on short-timescale variations ($\lesssim 100$ Myr), which dominate the evolution of dwarf galaxies in simulations \citep{Shen2014, Dominguez2015, Tacchella2022} and in observations of nearby galaxies \citep{Kauffmann2014, Weisz2014, Benitez2015}.  Burstiness also affects observed correlations such as the extensively-studied SFR-M$_*$ correlation \citep{Daddi2007, Noeske2007, Salim2007, Wuyts2011, Kurczynski2016} by adding scatter.  In particular, increased scatter has been predicted in simulations for dwarf galaxies \citep{Somerville2015}.  Additionally, \cite{Faucher-Giguere2018} has predicted a transition from ubiquitous bursty star formation in galaxies at high redshift to bursty star formation in low-mass galaxies and quiescent star formation in high-mass galaxies at low redshift.

Direct measurements of burstiness have been the subject of multiple recent papers for both observed \citep{Shen2014, Dominguez2015, Guo2016, Sparre2017, Broussard2019, Wang2020, Emami2021} and simulated \citep{Hopkins2014, Sparre2015} galaxies.  Many of these use the H$\alpha$ nebular emission line flux and the stellar ultraviolet continuum as tracers of recent star formation due to their sensitivity to stars formed in the past $\sim5$ Myr and $\sim200$ Myr respectively.  \cite{Weisz2012} took the ratio of these two fluxes, finding that a toy model of periodically repeating star formation bursts produced flux ratio distributions consistent with those of 185 \textit{Spitzer}-observed galaxies.  \cite{Guo2016} found a decrease in the average H$\beta$-to-far-UV SFR ratio with decreasing M$_*$, noting burstiness as a plausible explanation for this variation and that, with decreasing M$_*$, burstiness is increasingly necessary to reproduce the observed level of variation.  \cite{Wang2020} measured the power spectrum distribution (PSD) of star formation variability on various timescales, finding a power law slopes between 1.0 and 2.0 as well as a close relationship between specific star formation rate (sSFR) PSDs and the effective gas depletion timescale.

\cite{Broussard2019} showed that it is possible to estimate burstiness in a way that is robust to variations in the high-mass slope of the stellar IMF, metallicity, and (when assuming a \citealt{Calzetti2000} dust law) dust measurement errors for 3D-HST galaxies at $z\sim1$.  Because it is difficult to measure recent variations in a single galaxy's star formation history, burstiness is instead measured by comparing short- ($\mathbf{\lesssim10}$ Myr) and long-timescale ($\mathbf{\sim 100}$ Myr) star formation tracers using a burst indicator $\eta = \log_{10}(\mathrm{SFR_{H\alpha}}/\mathrm{SFR_{NUV}})$, which yields positive values in the case of rising star formation rates, negative values for falling star formation rates, and $\sim0$ for roughly constant star formation rates.  Consequently, the distribution of burst indicators for an ensemble of galaxies is used to characterize the population burstiness, with the width of the distribution being the primary indicator of bursty star formation.

Despite efforts to account for various survey selection effects when generating mock catalogs from simulations, \cite{Broussard2019} noted a discrepancy in the relationship between $\eta$ and M$_*$ when comparing the mock catalogs and 3D-HST observations.  The observed sample showed a strong positive trend for $\eta$ with increasing M$_*$ that was not present in the mocks.  Tuning the ratio of stellar to nebular dust reddening Q$_\mathrm{sg} = \mathrm{E(B-V)_{stars}/E(B-V)_{gas}}$ was only able to partially resolve the discrepancy, as modifying $Q_\mathrm{sg}$ from its typical low-redshift value of 0.44 \citep{Calzetti2000} to $\sim 1$ removed the correlation of $\eta$ with M$_*$, but also necessitated a negative average value of $\eta$.  This would have the unlikely implication of strongly declining SFHs for the selected sample of star forming galaxies. This motivates an independent treatment of the nebular and stellar dust in order to better understand burstiness at $z \gtrsim 1$.

While variations in Q$_\mathrm{sg}$ add systematic scatter to the distribution of galaxy $\eta$ values and therefore bias measurements of burstiness, previous burstiness analyses have not accounted for this effect.  In part, this is due to the rarity of observations that enable simultaneous independent dust measurements for each of the star formation tracers; however the relationship between $\mathrm{E(B-V)_{stars}}$ and $\mathrm{E(B-V)_{gas}}$ has been studied for a variety of dust curves outside of the context of burstiness \citep{Calzetti2000, Kashino2013, Price2014, Reddy2015, Shivaei2016, Reddy2020, Shivaei2020}.  This work presents the first analysis to apply measurements of the $Q_{sg}$ distribution to burstiness analyses.  Section \ref{Sec: Data} introduces the three surveys and two simulations we incorporate into our analysis.  Section \ref{Sec: Qsg} describes the sample selection and methodology of measuring $Q_{sg}$.  Our analysis consists of two approaches to measuring burstiness.  The first, described in Section \ref{Sec: Mocks} involves the generation of mock catalogs from the {\sc Mufasa} cosmological hydrodynamical simulation and Santa Cruz SAM that imitate the observational uncertainties and selection effects of our three observed surveys and therefore compare scatter in simulations against observations.  The second, described in Section \ref{Sec: eta_scatter} introduces a novel burstiness estimation technique in which we subtract away variance in $\eta$ caused by observational uncertainties and $Q_{sg}$ variations to measure the underlying intrinsic $\eta$ scatter.  These are followed by Section \ref{Sec: Results}, in which we describe the results of our two-pronged analysis.  We then discuss our findings in detail in Section \ref{Sec: Discussion} and present our final conclusions in Section \ref{Sec: Conclusions}.

\begin{figure}
    \centering
    \includegraphics[width=\columnwidth]{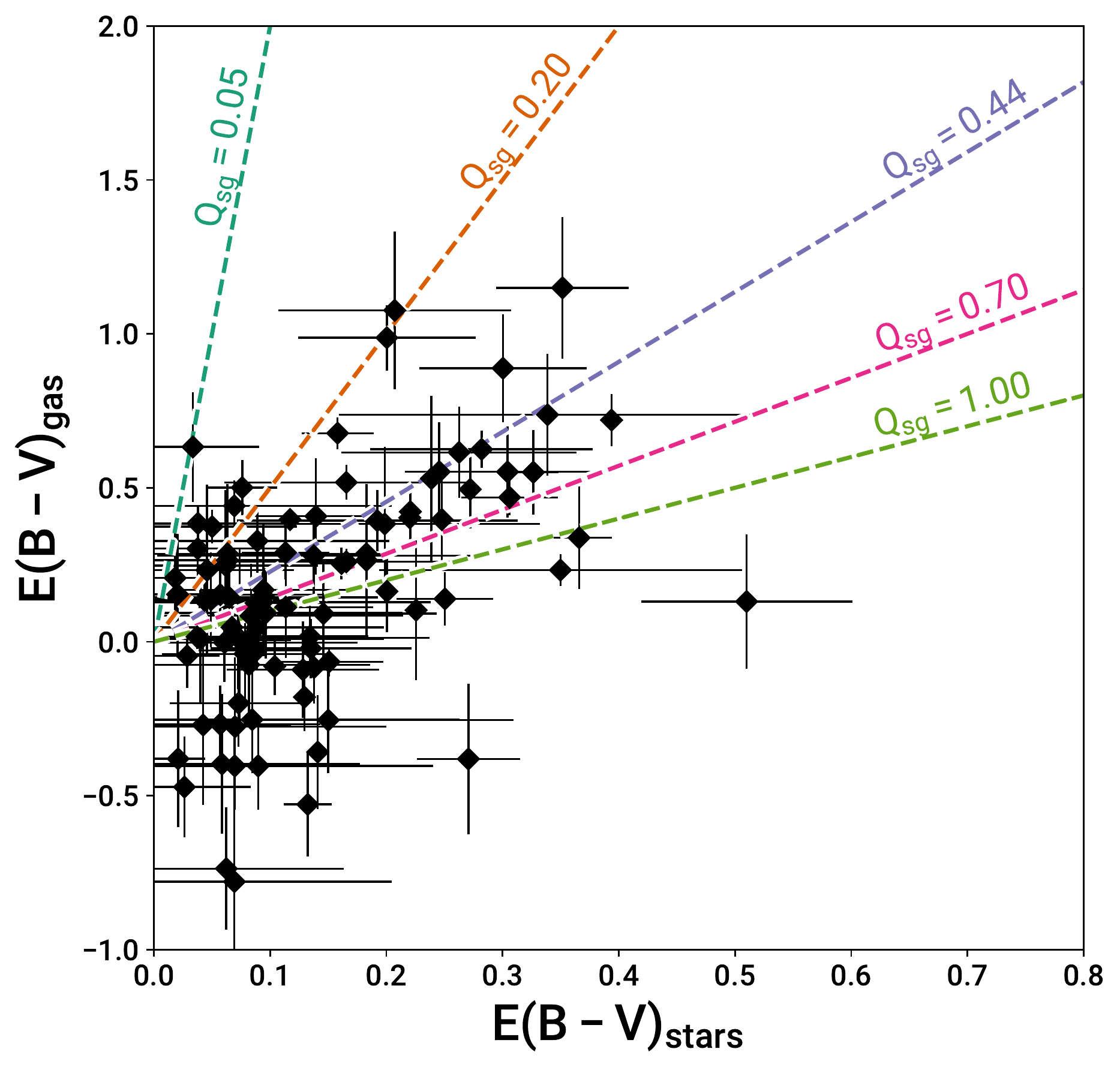}
    \caption{Scatter plot of $\mathrm{E(B-V)_{gas}}$ vs. $\mathrm{E(B-V)_{stars}}$ for MOSDEF galaxies assuming a \cite{Calzetti2000} dust law.  Dashed lines indicate various values of Q$_\mathrm{sg} = \mathrm{E(B-V)_{stars}}/\mathrm{E(B-V)_{gas}}$.  Many objects have $\mathrm{E(B-V)_{gas}} < 0$, caused by emission line ratios with $F_\mathrm{H\alpha}/F_\mathrm{H\beta} < 2.86$.  This indicates an unphysical inversion of the typical Balmer decrement dust correction.}
    \label{fig:EBV_comparison}
\end{figure}

\section{Data}\label{Sec: Data}
In addition to the 3D-HST \citep{Skelton2014} data set described in \cite{Broussard2019}, we incorporate the FMOS-COSMOS and MOSDEF surveys as described below.

\subsection{3D-HST Observations}

While we include a brief summary of the selections we implement for the 3D-HST catalog here to form the sample used throughout this work, a detailed description can be found in \cite{Broussard2019}.  This sample is selected to have $\mathrm{SNR_{H\alpha}} > 10$ to ensure acceptable data quality as determined by visual inspection of grism data for representative galaxies in the catalog.  We also restrict the redshift range to $0.65 < z < 1.50$ for consistency with detections of H$\alpha$ using the G141 grism.  Following \cite{Broussard2019}, we estimate an AGN contamination rate of $\sim3\%$ for our sample, which does not greatly affect our results, particularly because galaxies rejected for AGN contamination were not outliers in mass or SFR.  A constant [NII] fraction of 0.1 was assumed for all galaxies when calculating H$\alpha$ fluxes, and the dust correction was calculated from SED fits using a \cite{Calzetti2000} dust curve, which uses $\mathrm{E(B-V)_{stars}} = 0.44~E(B-V)_{gas}$.  3D-HST grism data were obtained by using a detection in the F 140W as a direct image, which serves to establish the calibration wavelength of the spectra. Detections in the G141 grism have an average 5$\sigma$ sensitivity of $f = 5.5 \times 10^{-17}~\mathrm{erg~s^{-1}~cm^{-2}}$ \citep{Brammer2012a}.

\subsection{FMOS-COSMOS Observations}

A second source of spectroscopic data comes from the FMOS-COSMOS survey \citep{Silverman2015}.  The Fiber Multi-Object Spectrograph (FMOS) is an instrument on the Subaru telescope in Mauna Kea, Hawaii.  The purpose of the FMOS-COSMOS survey is to provide high-resolution spectra of $\sim1000$ star forming galaxies at $z\sim 1.5$.  The H-long grating covers the wavelength range $1.6 < \lambda < 1.8 \mu \mathrm{m}$ enabling detections of H$\alpha$ while followup observations with the J-long grating ($ 1.11 < \lambda < 1.35 \mu \mathrm{m}$) adds coverage of H$\alpha$ and [OIII]$\lambda 4959,5007$.  Integration times enable detections of emission line fluxes with $f = 4 \times 10^{-17}~\mathrm{erg~s^{-1}~cm^{-2}}$ at $\mathrm{SNR = 3}$.  In order to alleviate sample selection bias against dust-obscured galaxies, \textit{Herschel} detections were used to select galaxies that lie on and off the star forming sequence.  AGN are identified with X-ray confirmation using \textit{Chandra} \citep{Civano2012} and excluded from our final sample.

\subsection{MOSDEF Observations}

We also use spectroscopy from the MOSFIRE Deep Evolution Field (MOSDEF; \cite{Kriek2015}) survey.  MOSDEF is a 47-night survey undertaken on the Keck MOSFIRE Spectrograph \citep{McLean2010, McLean2012} that covers $\sim 300$ sq. arcmin. across the AEGIS \citep{Davis2007}, COSMOS \citep{Scoville2007}, and GOODS-N \citep{Giavalisco2004} fields for the low-redshift regime of the survey ($1.37 < z < 1.70$), which is most comparable to 3D-HST.  MOSDEF excludes AGN using infrared, X-ray, and rest-frame optical line flux criteria described in \cite{Coil2015}, \cite{Azadi2017, Azadi2018}, and \cite{Leung2019}.  The public MOSDEF emission line catalog contains 3$\sigma$ emission line detections at a limiting line flux of $\sim 1\times 10^{17}~\mathrm{erg~s^{-1}~cm^{-2}}$.  We distinguish between two samples of MOSDEF galaxies in this work.  The first, which we will refer to as MOSDEF Gold, is a sample of 18 galaxies with simultaneous H$\alpha$ and H$\beta$ detections used in Section \ref{Sec: Qsg} to measure the central tendency and distribution width of the dust attenuation ratio $Q_{sg}$.  The second, which we refer to more generally as the MOSDEF sample, is the set of 263 MOSDEF galaxies with H$\alpha$ detections.  These are used to compare against the 3D-HST and FMOS-COSMOS samples using an assumed $Q_{sg}=0.43$ based on the results detailed in Section \ref{Sec: Qsg}.

All three spectroscopic surveys are supplemented with Gaussian Process SED fits \citep{Iyer2019} to photometry from CANDELS fields.  This flexible, non-parametric SFH reconstruction method produces star formation rates, stellar masses, and lookback times corresponding to quantiles of the galaxy's observed stellar mass.  Star formation histories are reconstructed using a brute-force Bayesian approach with a large pregrid of model SEDs, producing smooth star formation histories that are not reliant on any particular functional form.  This method also produces measurements of the dust attenuation and other standard galaxy physical properties.  We apply a $S/N > 1$ cut for the H$\alpha$ and SED measured SFRs for all surveys to exclude noise-dominated objects.

\subsection{Simulations}

We utilize two large-volume cosmological simulations to compare directly against observations: publicly available runs for {\sc Mufasa} \citep{Dave2017} and the Santa Cruz Semi-Analytic Model \citep[SAM;][]{Somerville2008, Somerville2015, Yung2019, Brennan2017}.  The {\sc Mufasa} hydrodynamical simulation attempts to directly model the fluid dynamics of the interstellar medium (ISM), intergalactic medium (IGM), and intercluster medium (ICM) in and around galaxies.  The Santa Cruz SAM instead achieves greater computational efficiency by applying physically motivated recipes to galaxies in dark matter halos computed from N-body simulations to determine their physical properties \citep{Somerville2015}.  Star formation histories are calculated using the Schmidt-Kennicutt law \citep{Kennicutt1998,Kennicutt1989} for normal quiescent star formation in isolated disks while merger-driven bursty star formation is based on recipes derived from hydrodynamical simulations of galaxy mergers in \cite{Robertson2006} and \cite{Hopkins2009}.

\subsection{Generating Tracer Star Formation Rates}

The process by which we generate tracer star formation rates is described in detail in \cite{Broussard2019}, however we include a brief summary here.  Calculations of the detailed time-response of various star formation tracers such as the H$\alpha$ emission line and NUV flux density is made possible with the use of Flexible Stellar Population Synthesis \citep[FSPS;][]{Conroy2009, Conroy2010}, a software package designed to generate realistic spectra of stellar populations from an input simulated SFH and physical properties such as dust attenuation, stellar initial mass function (IMF), and metallicity.  FSPS combines calculations of stellar evolution with stellar spectral libraries to produce spectra for simple stellar populations.  FSPS implements CLOUDY \citep{Byler2017} for calculating nebular emission.  CLOUDY is a photo-ionization code that assumes a constant density spherical shell of gas surrounding the stellar population.  Throughout this work, we use the input parameters of FSPS to specify a Chabrier IMF \citep{Chabrier2003}, include nebular line and continuum emission, implement a Calzetti dust law \citep{Calzetti2000}, and assume a gas-phase metallicity of $Z = 0.2 Z_\odot$.

\begin{figure}
    \centering
    \includegraphics[width=\linewidth]{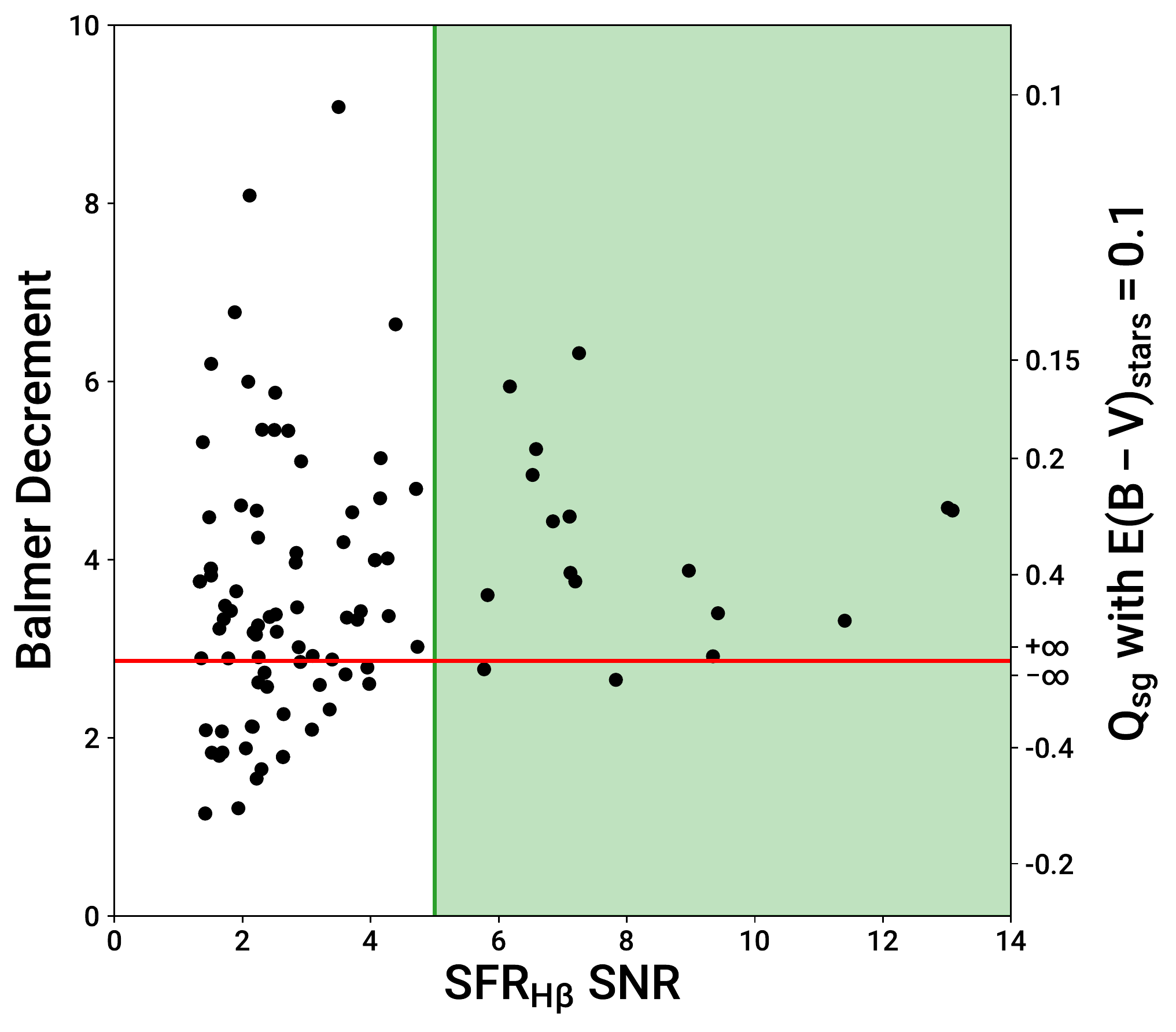}
    \caption{Balmer decrement ($F_\mathrm{H\alpha,uncorr}/F_\mathrm{H\beta,uncorr}$) and corresponding $Q_{sg}(\mathrm{assuming}~\mathrm{E(B-V)_{stars}}=0.1)$ vs. H$\beta$ SFR signal-to-noise ratio for MOSDEF galaxies with simultaneous H$\alpha$ and H$\beta$ detections.  The red horizontal line indicates a Balmer decrement of 2.86, which is expected in the absence of dust.  The green shaded region indicates $\mathrm{SFR_{H\beta}~SNR > 5}$, which we use to select the MOSDEF Gold sample used to measure the distribution of $Q_{sg}$.
    \label{fig:hb_snr_balmer_decrement}.}
\end{figure}

\begin{figure*}
    \centering
    \includegraphics[width = \textwidth]{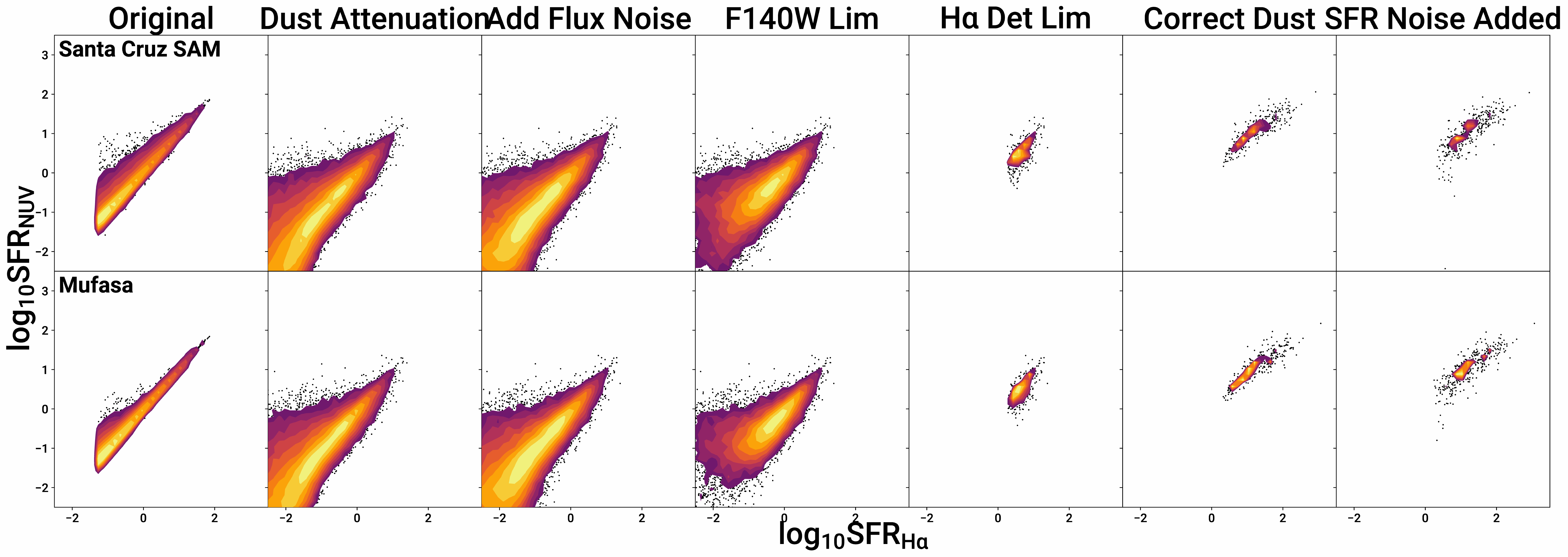}
    \caption{Plots showing the evolution of the Santa Cruz SAM (top row) and {\sc Mufasa} (bottom row) mock catalogs based on 3D-HST observations.  The detailed descriptions of each step involved in creating the mock catalog are described in Section \ref{Sec: Mocks}.}
    \label{fig:sfr_progression_3dhst}
\end{figure*}

\section{Measuring Variations in $Q_{sg}$} \label{Sec: Qsg}

The presence of two hydrogen lines for 111 galaxies in the MOSDEF survey enables direct dust correction of the H$\alpha$ flux via the Balmer decrement under the assumption of Case B recombination.  Combined with UV dust attenuation measurements from SED fitting, we have the means to constrain the detailed distribution of $Q_{sg} = \mathrm{E(B-V)_{stars}/E(B-V)_{gas}}$ values that represent the ratio of dust reddening between stellar and nebular light.

Solving for dust reddening under the assumption of Case B recombination yields the equation for converting from Balmer decrement to $\mathrm{E(B-V)_{gas}}$:
\begin{equation}
    \mathrm{E(B-V)_{gas}} = \log_{10}\left( \frac{F_\mathrm{H\alpha}}{2.86F_\mathrm{H\beta}} \right) \times 2.5 \left( k_\mathrm{H\beta} - k_\mathrm{H\alpha} \right)^{-1}
\end{equation}
where $F_\mathrm{H\alpha}$ and $F_\mathrm{H\beta}$ are the uncorrected emission line fluxes, and $k_\mathrm{H\alpha}$ and $k_\mathrm{H\beta}$ are the values of the assumed dust attenuation curve at the H$\alpha$ ($\lambda = 6563 \mathrm{\AA}$) and H$\beta$ ($\lambda = 4863 \mathrm{\AA}$) wavelengths respectively.

In practice, we find that 31 of the starting 111 galaxies exhibit Balmer ratios in which $F_\mathrm{H\alpha}/F_\mathrm{H\beta} < 2.86$, as we show in Figure \ref{fig:hb_snr_balmer_decrement}.  Similarly puzzling Balmer ratios are discussed in \cite{Boogaard2018} as being found in other spectroscopic surveys and have several possible explanations, but in our case, these objects typically have high relative uncertainty and are likely the result of scatter.  As a result, we limit the sample used to calculate the $Q_{sg}$ distribution to only those galaxies with $\mathrm{SFR_{H\beta}}/\sigma_\mathrm{SFR_{H\beta}} > 5$, eliminating all but two objects with such Balmer decrements, and leaving 18 galaxies in total (the MOSDEF Gold sample).  From these 18 galaxies, we find $\operatorname{med}(Q)_{sg}=0.43$, consistent with dust measurements in low-redshift star forming galaxies \citep{Calzetti2000} and the width of the $Q_{sg}$ distribution $\sigma_{Q_{sg}} = 0.33$, calculated via a robust estimator.

Here, it is also important to note that because $Q_{sg} = \mathrm{E(B-V)_{stars}}/\mathrm{E(B-V)_{gas}}$, Balmer decrements greater than 2.86 yield $\mathrm{E(B-V)_{gas}} > 0$ (and therefore $Q_{sg} > 0$) while Balmer decrements less than 2.86 yield $\mathrm{E(B-V)_{gas}} < 0$ (and therefore $Q_{sg} < 0$).  Further, a Balmer decrement of $\sim2.86$ indicates a galaxy with nearly zero nebular dust reddening, meaning that $Q_{sg}$ tends toward negative infinity as the Balmer decrement tends toward 2.86 from below, while $Q_{sg}$ tends toward positive infinity as the Balmer decrement tends toward 2.86 from above, as is indicated by the right axis in Figure \ref{fig:hb_snr_balmer_decrement}.

\section{Mock Catalog Generation} \label{Sec: Mocks}

In order to better understand the predictions of the SAM and {\sc Mufasa} simulations, we follow \cite{Broussard2019} by creating a mock catalog from each simulation that is designed to mimic the observational systematics of each observed data set.  Throughout the mock catalog generation process, each time bin of a simulated galaxy's SFR is treated individually, meaning that a single galaxy's star formation history can contribute multiple ``independent'' SFR values to the mock catalog, and for the purpose of the mock catalog, each of these is treated as a separate galaxy (though we will later restrict the sample to be within $|\Delta z| < 0.2$ of each survey's mean redshift).  While a description of each step in the mock catalog generation process for each of the observed samples is detailed below, a visual summary is shown in Figure \ref{fig:sfr_progression_3dhst} for 3D-HST, Figure \ref{fig:sfr_progression_fmoscosmos} for FMOS-COSMOS, and Figure \ref{fig:sfr_progression_mosdef} for the MOSDEF sample, assuming $Q_{sg}=0.43$ for all three.  These summary figures demonstrate the progression in $\mathrm{SFR_{NUV}}$ vs. $\mathrm{SFR_{H\alpha}}$ as each selection effect is modeled.  For clarity, throughout the remainder of this work we will refer to the addition of the effects of dust to simulated fluxes as dust attenuation while the removal of the effects of dust from simulated or observed fluxes will be referred to as dust correction.

\subsection{Initial SFR$_{H\alpha}$ Draws} \label{Subsec: sfrha_draws}

To form our initial sample of galaxies for the mock catalog, we begin by drawing SFR measurements such that our starting galaxy sample matches the observed star formation rate function (SFRF) from the High Redshift (Z) Emission Line Survey (HIZELS; \citealt{Geach2008, Sobral2012, Sobral2013}).  HIZELS is a narrowband survey and consequently, star formation rate functions are available as Schechter function fits at $z = 0.4,~0.84,~1.47,~\mathrm{and}~2.23$.  Because the 3D-HST sample has an average redshift of $\Bar{z}\approx 1$, we interpolate the SFRF parameters between $z = 0.84$ and $z = 1.47$ to approximate the SFRF at $z\sim 1$.  We perform a similar interpolation for FMOS-COSMOS using the average redshift of $z\sim 1.59$ and for MOSDEF using $z\sim1.54$.  The resulting Schechter function parameters are summarized in Table \ref{Tbl: Schechter_params}.

\begin{deluxetable*}{c|cccc}
\tablecaption{Schechter Function Parameters}
\tablehead{ \colhead{Sample} & \colhead{Avg. Redshift} & \colhead{SFR$^*_{\alpha = -1.6}$} & \colhead{$\log_{10}(\Phi^*_{\alpha = -1.6})$} & \colhead{$\alpha$}\\[-.2cm] \colhead{} & \colhead{($\Bar{z}$)} & \colhead{(M$_\odot$ yr$^{-1}$)} & \colhead{(Mpc$^{-3}$)} & \colhead{}}
\startdata
3D-HST & 1.00 & 13.98 & -2.67 & -1.65 \\
FMOS-COSMOS & 1.59 & 30.89 & -2.75 & -1.61 \\
MOSDEF ($1.37<z<1.7$) & 1.54 & 28.23 & -2.73 & -1.59 \\
\enddata
\tablecomments{\label{Tbl: Schechter_params}A table of the average redshifts for each survey as well as the interpolated Schechter parameters used to select the initial sample of mock galaxies based on their SFR$_\mathrm{H\alpha}$ described in Section \ref{Subsec: sfrha_draws}.}
\end{deluxetable*}

\begin{figure*}
    \centering
    \includegraphics[width=\textwidth]{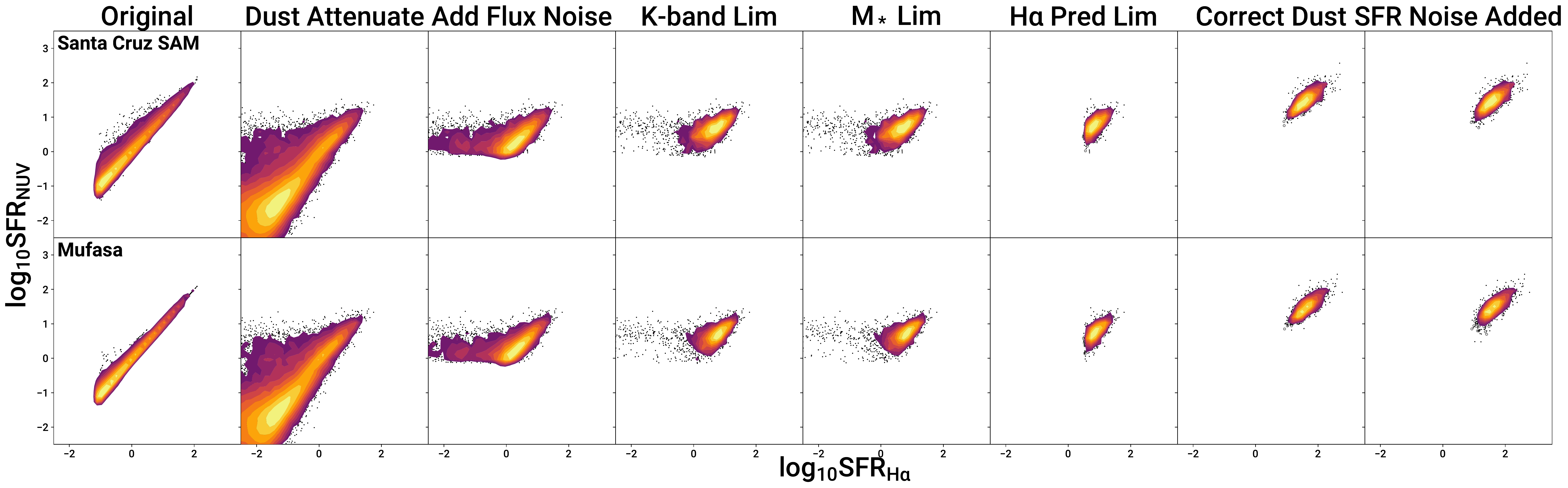}
    \caption{Plots showing the evolution of the Santa Cruz SAM (top row) and {\sc Mufasa} (bottom row) mock catalogs based on FMOS-COSMOS observations.  The detailed descriptions of each step involved in creating the mock catalog are described in Section \ref{Sec: Mocks}.}
    \label{fig:sfr_progression_fmoscosmos}
\end{figure*}

\begin{figure*}
    \centering
    \includegraphics[width = \textwidth]{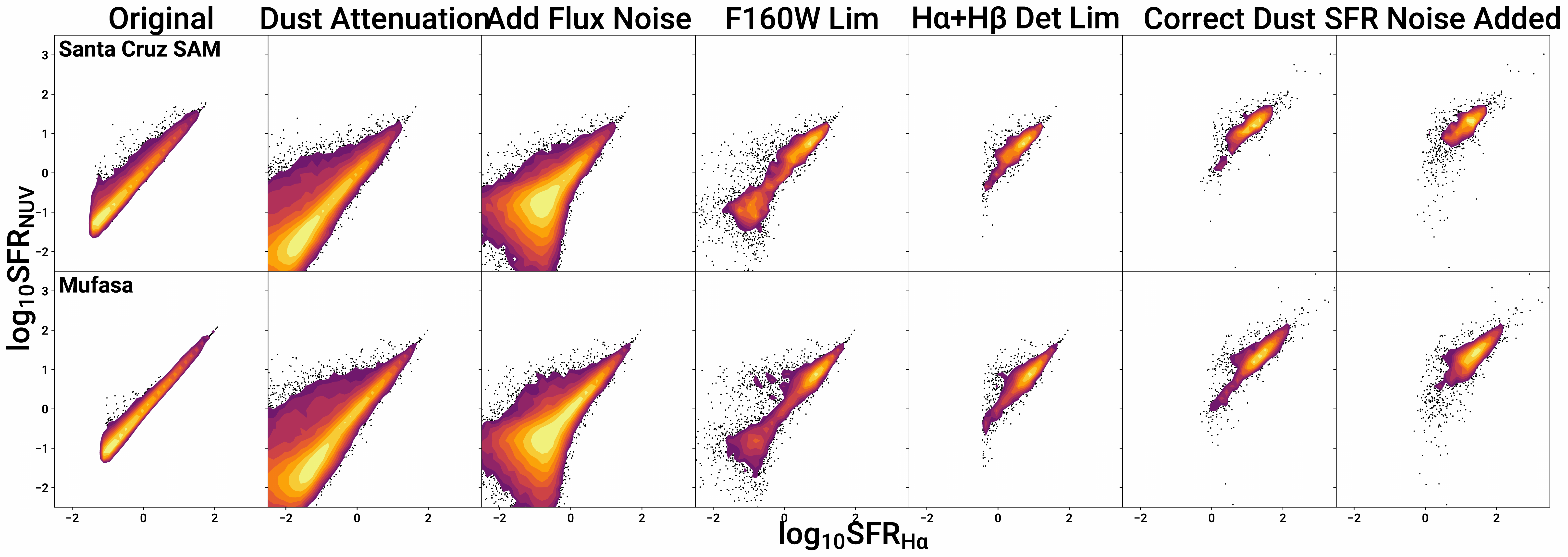}
    \caption{Plots showing the evolution of the Santa Cruz SAM (top row) and {\sc Mufasa} (bottom row) mock catalogs based on MOSDEF observations.  The detailed descriptions of each step involved in creating the mock catalog are described in Section \ref{Sec: Mocks}.}
    \label{fig:sfr_progression_mosdef}
\end{figure*}

We draw galaxies from the SAM and {\sc Mufasa} simulations by normalizing the SFRF to 1 by integrating over $10^{-3} < \mathrm{SFR/SFR^*} < 10^2$, effectively producing a probability distribution function (PDF) such that $p(\mathrm{SFR}) \propto \phi(\mathrm{SFR})$.  When drawing SFRs from this distribution, we select an SFR from the SAM or {\sc Mufasa} simulation that lies within 0.5 dex of the drawn SFR and with $|z_\mathrm{sim} - \Bar{z}_\mathrm{obs}| < 0.2$ where $z_\mathrm{sim}$ is the redshift of the matched SAM or {\sc Mufasa} SFR and $\Bar{z}_\mathrm{obs}$ is the average redshift of the particular mock's corresponding survey.  Finally, SFR$_\mathrm{H\alpha}$, SFR$_\mathrm{NUV}$, and $M_*$ for the matched galaxy are all scaled such that SFR$_\mathrm{H\alpha}$ exactly matches the original SFRF-drawn SFR.  This process is repeated for $10^5$ galaxy draws from this distribution for each of the mock catalogs, producing the initial distribution of galaxies shown in the leftmost panels of Figures \ref{fig:sfr_progression_3dhst}, \ref{fig:sfr_progression_fmoscosmos}, and \ref{fig:sfr_progression_mosdef}.

\subsection{Dust Attenuation}

We use FSPS to derive the relationship between star formation rate and the observed H$\alpha$ luminosity (L$_\mathrm{H\alpha}$) and NUV luminosity density (L$_\mathrm{\nu,NUV}$) by generating dust-free galaxy spectra for a constant star formation rate, finding that $\nu_\mathrm{NUV} \mathrm{L}_\mathrm{\nu,NUV} = \mathrm{SFR_{NUV} \times 10^{42.94}~ \mathrm{erg~s^{-1}~Hz^{-1}}} / (\mathrm{M_\odot~yr^{-1}})$ and $\mathrm{L_{H\alpha}} = \mathrm{SFR_{H\alpha} \times 10^{41.20}~erg~s^{-1}/(M_\odot~yr^{-1})}$.  Using these relationships, we convert the mock SFR$_\mathrm{H\alpha}$ and SFR$_\mathrm{NUV}$ into their corresponding flux and flux density respectively so that dust attenuation can be applied.

For each observed sample, we perform fits of various galaxy properties against $\mathrm{A_V}$ to determine the most robust relationship.  For 3D-HST, we find a tight correlation between SFR$_\mathrm{H\alpha}$ and $\mathrm{A_V}$.  For FMOS-COSMOS, we find the strongest correlation to be between $\mathrm{A_V}$ and SFR$_\mathrm{NUV}$.  Meanwhile, for MOSDEF, we find the strongest correlation to be between $\mathrm{A_V}$ and $\mathrm{M_*}$.  For each of these cases, we fit a Gaussian mixture model (GMM) to each relationship, enabling random assignment of $\mathrm{A_V}$ values to mock galaxies in a way that preserves the covariance between $\mathrm{A_V}$ and the best-correlated galaxy property.

Because of the addition of the MOSDEF sample to this analysis that includes simultaneous detections of the H$\alpha$ and H$\beta$ emission lines, we are able to characterize the distribution of $Q_\mathrm{sg}$ values as shown in Figure \ref{fig:hb_snr_balmer_decrement}.  We assign values of $Q_\mathrm{sg}$ (which we call $Q_\mathrm{sg,intrinsic}$) to drawn galaxies using a Gaussian distribution with a $\mu = 0.43$ and $\sigma = 0.33$.  We set a minimum assignable value of 0.05 to avoid negative values of $Q_\mathrm{sg,intrinsic}$, which would imply an unphysical, negative $\mathrm{E(B-V)_{gas}}$.

\subsection{Flux Uncertainties}

We approximate the uncertainty of a given simulated galaxy's H$\alpha$ flux by first fitting the relationship between the H$\alpha$ flux its associated uncertainty for each of the three observed samples.  This process is described in detail in \cite{Broussard2019}, and we perform a similar fit here to each of the 3D-HST, FMOS-COSMOS, and MOSDEF samples to assign realistic flux uncertainties to mock galaxy H$\alpha$ fluxes and NUV flux densities.


\subsection{Flux Limits}

In order to model the various observational limits and selection effects of each survey, we use FSPS to directly estimate the flux density in detection bands using the detailed star formation history and assigned $A_V$ value for each mock galaxy.  The detailed selection criteria are listed below for each survey.

The 3D-HST survey requires a detection in the F140W filter, which is used as the calibration image for the grism data.  This corresponds to a 25.8 magnitude limit in the F140W band.  We further exclude all mock galaxies that do not meet the $\mathrm{SNR > 8}$ grism detection limit of $8.8 \times 10^{-17}~\mathrm{erg~s^{-1}~cm^{-2}}$.

The FMOS-COSMOS sample uses a series of sample selections.  A cut in $(B -z) - (z-K)$ selects star forming galaxies based on the findings of \cite{Daddi2004}.  A cut in K-band magnitude of $K_s < 23$ is based on the depth and completeness of CFHT WIRCam observations \citep{McCracken2010}. Targets are also excluded if the estimated H$\alpha$ flux is below the 3$\sigma$ detection limit of $4\times10^{-17}~\mathrm{~erg~s^{-1}~cm^{-2}}$, with the flux estimated by treating the UV SFR as an H$\alpha$ SFR and combining it with the spectroscopic redshift to yield an observed line flux.  Additionally, a mass cut of $log_{10}(M_*) > 9.8$ is implemented to our mock galaxies to match the targeted mass range of the survey.

MOSDEF implements a magnitude cut of $H < 24$ for the low-redshift interval $1.37<z<1.70$ when selecting targets.  Although the initial pool of targets is itself composed of objects with detections in the 3D-HST photometric catalog \citep{Skelton2014}, we do not implement the combined $F125W + F140W + F160W$ image detection criteria as \cite{Skelton2014} notes that the sample is $\sim 90\%$ complete even in the shallow CANDELS fields when selecting only objects with $H < 25$.  We also exclude mock objects that are not sufficiently bright to meet the $SNR > 3$ MOSDEF detection limit of $\sim1.7\times10^{-17}~\mathrm{~erg~s^{-1}~cm^{-2}}$.

\subsection{Dust Correction}

With the observed systematics implemented, we correct for dust in each mock galaxy.  Each mock catalog corrects for dust using a ``measured'' value of $\mathrm{A_V}$ that has scatter added based on the relationship between $\mathrm{A_V}$ and $\sigma_\mathrm{A_V}$, assuming Q$_\mathrm{sg,measured}=  0.43$ across all simulated observations.  As a result, the effect of adding dust attenuation to the mocks and correcting it in this later step for galaxies that are not excluded by survey selection effects is an increase in flux scatter relative to the intrinsic values due to the associated $A_V$ uncertainty and measurement errors that arise from assuming a single value of $Q_{sg,assumed}=0.43$.  For reference $Q_{sg} > Q_{sg,\mathrm{assumed}}$ will yield $F_\mathrm{H\alpha,meas} > F_\mathrm{H\alpha}$ for the same $A_V$ (and vice-versa).

\subsection{Matching SFR Measurement Uncertainties}

To ensure that the mock catalogs are generating systematic scatter that is comparable to that of each corresponding observed sample, we track each galaxy through the mock catalog generation process.  After applying each of the previous steps, we compute the difference between each remaining galaxy's resulting SFR$_\mathrm{H\alpha}$ (SFR$_\mathrm{NUV}$) and the SFR$_\mathrm{H\alpha}$ (SFR$_\mathrm{NUV}$) it was assigned during the initial draw in Section \ref{Subsec: sfrha_draws}, producing an effective mock observational error.  We then bin galaxies by SFR$_\mathrm{H\alpha}$ and SFR$_\mathrm{NUV}$, computing the average uncertainty of the observed galaxy in each bin and the standard deviation of the mock observational error and add Gaussian scatter to the relevant mock SFRs with $\sigma_\mathrm{SFR,added} = \sqrt{\sigma_\mathrm{SFR,obs}^2 - \sigma_\mathrm{SFR,mock}^2}$.  This causes the mock observational error in each bin to rise to match the average measurement uncertainty, thus ensuring that the mocks do not underestimate the scatter inherent in computing SFRs from observations.  In practice, this adds scatter to fewer than half of the bins, and the added scatter well under 1 dex.

\begin{figure*}
    \centering
    \includegraphics[width = \linewidth]{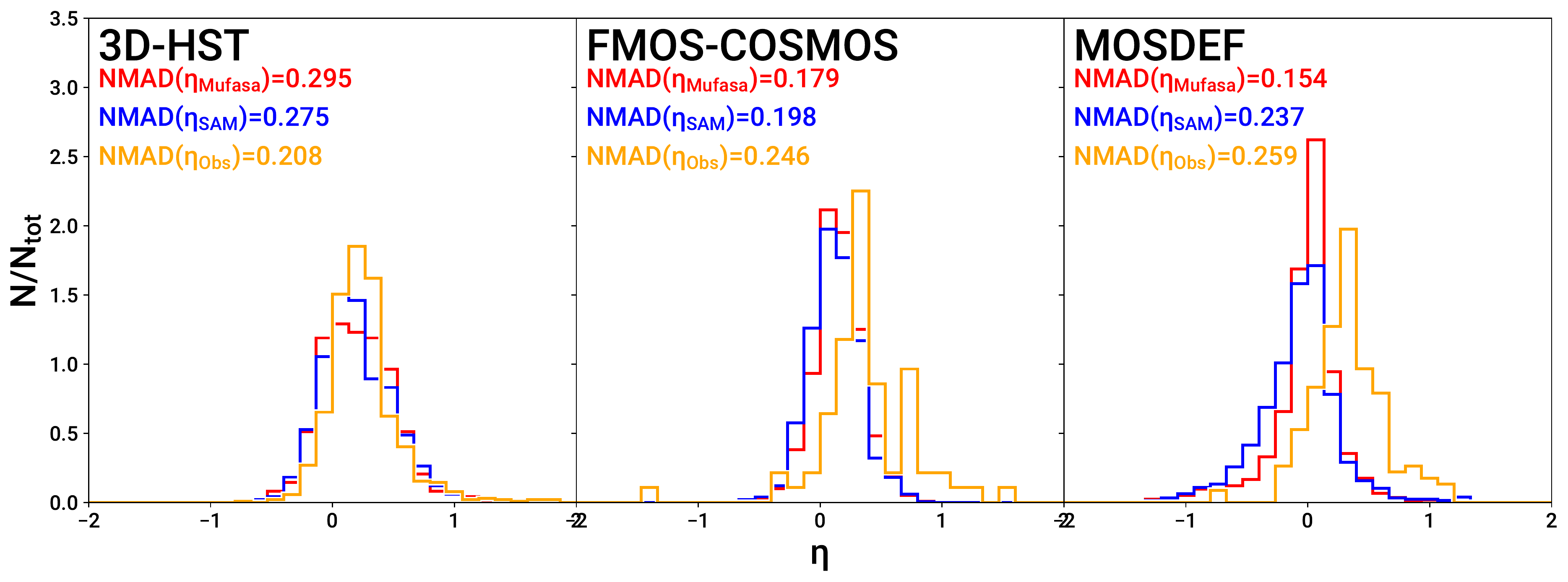}
    \caption{Distributions of $\eta$ for the {\sc Mufasa} mocks (red), SAM mocks (blue), and observed galaxies (yellow).  We see good agreement between distribution shapes and widths for the 3D-HST mock and observed samples, but the FMOS-COSMOS and MOSDEF mocks under-estimate the median value and scatter in $\eta$.}
    \label{fig:all_burstiness_hists}
\end{figure*}

\begin{deluxetable*}{c|crc}
\tablecaption{$\eta$ Distribution Statistics}
\tablehead{ \colhead{Survey} & \colhead{Sample} & \colhead{$\widetilde{\eta}$} & \colhead{$\operatorname{NMAD}(\eta)$}}
\startdata
 & Observed & 0.175 & 0.208 \\
3D-HST & SAM mock & 0.165 & 0.275 \\
 & {\sc Mufasa} mock & 0.193 & 0.295 \\ \hline
 & Observed & 0.302 & 0.246 \\
FMOS-COSMOS & SAM mock & 0.116 & 0.198 \\
 & {\sc Mufasa} mock & 0.146 & 0.179 \\ \hline
  & Observed & 0.054 & 0.259 \\
MOSDEF ($1.37<z<1.7$) & SAM mock & -0.070 & 0.235 \\
 & {\sc Mufasa} mock & 0.004 & 0.151 \\
\enddata
\tablecomments{\label{Tbl: eta_stats} Descriptive statistics of the various $\eta$ distributions shown in Figure \ref{fig:all_burstiness_hists}.  Here, we denote the median of $\eta$ as $\widetilde{\eta}$.}
\end{deluxetable*}

\section{Correcting for Systematic Scatter in $\eta$} \label{Sec: eta_scatter}

Thus far, we have endeavored to apply forward modeling techniques to simulated data in an effort to compare burstiness between the mock and observed samples.  Here, we will show that, with accurate knowledge of observational uncertainties, it is also possible to estimate the underlying burstiness of an observed sample, which can then be compared directly against the $\eta$ distribution for simulated galaxies (e.g., after making a reasonable cut in $M_*$).  

For a single galaxy, we can express the observed burst indicator value $\eta_\mathrm{obs}$ as the sum of some true underlying burst indicator value $\eta$ and the measurement error $\delta_\eta$:
\begin{align}
    \eta_\mathrm{obs} ={}& \eta + \delta_{\eta}.\label{Eqn: eta_obs}
\end{align}
Shifting now to a view of a population of such galaxies, we can describe the variance in $\eta_\mathrm{obs}$ in terms of the intrinsic $\eta$ variance and the ``explainable'' portion of the observed variance for an ensemble of galaxies using Equation \ref{Eqn: eta_obs}:
\begin{align}
    \Var(\eta_\mathrm{obs}) = {}& \Var_\mathrm{int}(\eta) + \Var(\delta_\eta)\label{Eqn: var_eta_obs}\\ \nonumber
    = {}& \Var_\mathrm{int}(\eta) + \Var_\mathrm{exp}(\eta_\mathrm{obs}).
\end{align}

\subsection{Dominant SFR Uncertainties}
In the case where SFR uncertainties dominate the $\eta$ measurement uncertainty, it is possible to break $\eta$ into its constituent parts using its definition as the logarithm of  the ratio of two SFRs:
\begin{align}
    \eta_\mathrm{obs} ={}& \log_{10}\left( \frac{\mathrm{SFR_{H\alpha,obs}}}{\mathrm{SFR_{NUV,obs}}} \right). \label{Eqn: eta_true}
\end{align}
This yields the following equation for the explained variance in $\eta_\mathrm{obs}$ in terms of the SFR measurement errors:
\begin{align}
    \Var_\mathrm{exp}(\eta_\mathrm{obs}) = {}& \Var(\delta_{\log_{10}\mathrm{SFR_{H\alpha,obs}}}) \nonumber \\
    & + \Var(\delta_{\log_{10}\mathrm{SFR_{NUV,obs}}}) \nonumber \\
    & - 2 \Cov(\delta_{\log_{10}\mathrm{SFR_{H\alpha,obs}}},\delta_{\log_{10}\mathrm{SFR_{NUV,obs}}}).
\end{align}
Here, we set the final term to zero because the measurement errors for each star formation rate should be linearly independent, and we can rewrite the expression in terms of the galaxies' SFR uncertainties:
\begin{align}
    \Var_\mathrm{exp}(\eta_\mathrm{obs}) = {}& \Avg(\sigma^2_\mathrm{\log_{10}SFR_{H\alpha,obs}}) \nonumber \\
    & + \Avg(\sigma^2_\mathrm{\log_{10}SFR_{NUV,obs}}).
\end{align}
While this method turns out to be insufficient for describing the variance in $\eta$ because it does not take into account detailed dust systematics, it is a useful starting point for the more complex derivations to come.

\subsection{Flux and Dust Attenuation Uncertainties with Known Balmer Decrements}

We consider two differing regimes - galaxy samples with and without measured Balmer decrements - when calculating the scatter in $\eta_\mathrm{obs}$.  Starting with the case of known Balmer decrements, we can expand the two star formation rates into quantities that are either directly observed or inferred through redshift fitting and spectral energy distribution (SED) fitting yields the following sequence of equations, culminating in the redefinition of $\eta$ below in terms of the galaxy's observed flux (density), the SFR conversion factors, the chosen dust law, and the dust attenuation of light emitted by nebular and stellar light respectively.
\begin{align}
    \mathrm{SFR_{H\alpha}} ={}& C_\mathrm{H\alpha} (4\pi d_\mathrm{L}^2)F_\mathrm{H\alpha}\times 10^{0.4k_\mathrm{H\alpha}\mathrm{E(B-V)_{gas}}}\\
    \mathrm{SFR_{NUV}} ={}& C_\mathrm{NUV} (4\pi d_\mathrm{L}^2)\nu F_\mathrm{\nu,NUV}\times 10^{0.4k_\mathrm{NUV}\mathrm{E(B-V)_{stars}}}\\
    \frac{\mathrm{SFR_{H\alpha}}}{\mathrm{SFR_{NUV}}} ={}& \frac{C_\mathrm{H\alpha}F_\mathrm{H\alpha}}{C_\mathrm{NUV}\nu F_\mathrm{\nu,NUV}} \nonumber\\
    & \times 10^{0.4[k_\mathrm{H\alpha}\mathrm{E(B-V)_{gas}} - k_\mathrm{NUV}\mathrm{E(B-V)_{stars}}]}\\
    \eta={}& \log_{10}\left( \frac{C_\mathrm{H\alpha}F_\mathrm{H\alpha}}{C_\mathrm{NUV}\nu F_\mathrm{\nu,NUV}} \right) \nonumber\\
    & + 0.4\left[ k_\mathrm{H\alpha}\mathrm{E(B-V)_{gas}} - k_\mathrm{NUV}\mathrm{E(B-V)_{stars}} \right] \label{Eqn: eta_true_breakdown}
\end{align}

Under the assumption of a \cite{Calzetti2000} dust law ($k_\mathrm{H\alpha}$, $k_\mathrm{NUV}$) and \cite{Kennicutt2012} SFR conversion factors ($C_\mathrm{H\alpha}$, $C_\mathrm{NUV}$), we are left with four measured parameters with associated uncertainties: $\log_{10} F_\mathrm{H\alpha}$, $\log_{10}F_\mathrm{NUV}$, $\mathrm{E(B-V)_{gas}}$, and $\mathrm{E(B-V)_{stars}}$, 
where we have adopted $F_\mathrm{NUV}\equiv \nu F_\mathrm{\nu,NUV}$ to simplify the subsequent notation.  It is apparent by examining Equations \ref{Eqn: eta_obs} and \ref{Eqn: eta_true} that we can replace each of these four parameters with a term of the form $x + \delta_x$, representing the sum of its intrinsic value and the individual observational error. This resulting equation would then describe $\eta_\mathrm{obs}$ for the hypothetical galaxy, with the observational error terms summing to $\delta_\eta$.

Returning to Equation \ref{Eqn: eta_true_breakdown} we can express $\eta_\mathrm{obs}$ for a single galaxy in terms of true values and their observational errors:

\begin{align}
    \eta_\mathrm{obs} ={}& \log_{10} (C_\mathrm{H\alpha}) \nonumber\\
    & + \log_{10}(F_\mathrm{H\alpha}) + \delta_{\log_{10}F_\mathrm{H\alpha}} \nonumber\\
    & + 0.4k_\mathrm{H\alpha}\left[\mathrm{E(B-V)_{gas}} + \delta_\mathrm{E(B-V)_{gas}}\right] \nonumber\\
    & - \log_{10} (C_\mathrm{NUV}\nu) \nonumber\\
    & - \log_{10}(F_\mathrm{NUV}) - \delta_{\log_{10}F_\mathrm{NUV}} \nonumber\\
    & - 0.4k_\mathrm{NUV}\left[\mathrm{E(B-V)_{stars}} + \delta_\mathrm{E(B-V)_{stars}}\right].
\end{align}
We then continue by subtracting away the true burst indicator $\eta$ and taking the variance to isolate the explained variance from observations $\Var_\mathrm{exp}(\eta_\mathrm{obs})$ (i.e., the scatter in $\eta_\mathrm{obs}$ caused by observational uncertainties), which contains four variance terms and six covariance terms:
\begin{align}
    \Var_\mathrm{exp}(\eta_\mathrm{obs}) ={}& \Var(\delta_{\log_{10}F_\mathrm{H\alpha}}) \nonumber\\
    & + \Var(\delta_{\log_{10} F_\mathrm{NUV}}) \nonumber\\
    & + (0.4k_\mathrm{H\alpha})^2 \Var(\delta_{\mathrm{E(B-V)_{gas}}}) \nonumber\\
    & + (0.4k_\mathrm{NUV})^2 \Var(\delta_{\mathrm{E(B-V)_{stars}}}) \nonumber\\
    & - 2 \Cov(\delta_{\log_{10}F_\mathrm{H\alpha}},\delta_{\log_{10} F_\mathrm{NUV}}) \nonumber\\
    & + 2 (0.4 k_\mathrm{H\alpha}) \Cov(\delta_{\log_{10}F_\mathrm{H\alpha}}, \delta_{\mathrm{E(B-V)_{gas}}}) \nonumber\\
    & - 2 (0.4 k_\mathrm{NUV})\Cov(\delta_{\log_{10}F_\mathrm{H\alpha}}, \delta_{\mathrm{E(B-V)_{stars}}}) \nonumber\\
    & - 2 (0.4k_\mathrm{H\alpha}) \Cov(\delta_{\log_{10}F_\mathrm{NUV}}, \delta_{\mathrm{E(B-V)_{gas}}}) \nonumber\\
    & + 2 (0.4k_\mathrm{NUV}) \Cov(\delta_{\log_{10}F_\mathrm{NUV}}, \delta_{\mathrm{E(B-V)_{stars}}}) \nonumber\\
    & - 2 (0.16k_\mathrm{H\alpha}k_\mathrm{NUV}) \Cov(\delta_{\mathrm{E(B-V)_{stars}}}, \delta_{\mathrm{E(B-V)_{gas}}}).
\end{align}
The observational errors associated with measuring $F_\mathrm{H\alpha}$ and $F_\mathrm{NUV}$ are uncorrelated, as are the errors associated with each of the $\mathrm{E(B-V)}$ terms.  While there is possibly some covariance between $\delta_{\log_{10}F_\mathrm{NUV}}$ and $\delta_\mathrm{E(B-V)_{stars}}$ because of the dependence of dust measurements on UV emission, dust corrections derived using SED fitting (as is the case for each of our samples) are not strongly dependent on any single band.  As a result, we set each of these covariance terms to zero, leaving only the four variance terms:
\begin{align}
    \Var_\mathrm{exp}(\eta_\mathrm{obs}) ={}& \Var(\delta_{\log_{10}F_\mathrm{H\alpha}}) \nonumber\\
    & + \Var(\delta_{\log_{10}F_\mathrm{NUV}}) \nonumber\\
    & + (0.4k_\mathrm{H\alpha})^2 \Var(\delta_{\mathrm{E(B-V)_{gas}}}) \nonumber\\
    & + (0.4k_\mathrm{NUV})^2 \Var(\delta_{\mathrm{E(B-V)_{stars}}}). 
\end{align}
Finally, we can express the above equation in terms of the reported uncertainties to arrive at an expression for the scatter added to $\eta$ as a result of observational uncertainties:
\begin{align}
    \Var_\mathrm{exp}(\eta_\mathrm{obs}) ={}& \operatorname{Avg}(\sigma^2_{\log_{10}F_\mathrm{H\alpha}}) \nonumber\\
    & + \operatorname{Avg}(\sigma^2_{\log_{10}F_\mathrm{NUV}}) \nonumber\\
    & + (0.4k_\mathrm{H\alpha})^2 \operatorname{Avg}(\sigma^2_{\mathrm{E(B-V)_{gas}}}) \nonumber\\
    & + (0.4k_\mathrm{NUV})^2 \operatorname{Avg}(\sigma^2_{\mathrm{E(B-V)_{stars}}}).\label{Eqn: scatter_balmer} 
\end{align}

\subsection{Flux and Dust Attenuation Uncertainties without Balmer Decrements}

When Balmer Decrements are not available for an ensemble of galaxies, we can instead use the relationship $\mathrm{E(B-V)_{gas}} = \mathrm{E(B-V)_{stars}}/Q_\mathrm{sg}$ to redefine Equation \ref{Eqn: eta_true_breakdown} in terms of $F_\mathrm{H\alpha}$, $F_\mathrm{NUV}$, $\mathrm{E(B-V)_{stars}}$, and $Q_\mathrm{sg}$:
\begin{align}
    \eta ={}& \log_{10}\left( \frac{C_\mathrm{H\alpha}F_\mathrm{H\alpha}}{C_\mathrm{NUV}\nu F_\mathrm{NUV}} \right) \nonumber\\
    & + 0.4\mathrm{E(B-V)_{stars}}\left[ k_\mathrm{H\alpha}Q_\mathrm{sg}^{-1} - k_\mathrm{NUV} \right].
\end{align}
Continuing, this gives the following relationship for $\eta_\mathrm{obs}$.

\begin{align}
    \eta_\mathrm{obs} ={}& \log_{10}\left( \frac{C_\mathrm{H\alpha}}{\nu C_\mathrm{NUV}} \right) \nonumber\\
    & + \log_{10}F_\mathrm{H\alpha} + \delta_{\log_{10} F_\mathrm{H\alpha}}\nonumber\\
    & - \log_{10}F_\mathrm{NUV} - \delta_{\log_{10} F_\mathrm{NUV}}\nonumber\\
    & + 0.4(k_\mathrm{H\alpha}Q_{sg}^{-1} - k_\mathrm{NUV})\mathrm{E(B-V)_\mathrm{stars}}  \nonumber \\
    & + 0.4 (k_\mathrm{H\alpha}Q_{sg}^{-1}-k_\mathrm{NUV})\delta_\mathrm{E(B-V)_{stars}} \nonumber\\
    & + 0.4 k_\mathrm{H\alpha} \mathrm{E(B-V)_{stars}} \delta_{Q_{sg}^{-1}} \nonumber \\
    & + 0.4 k_\mathrm{H\alpha} \delta_\mathrm{E(B-V)_{stars}} \delta_{Q_{sg}^{-1}}
\end{align}



Similar to the case with known Balmer decrements, we subtract away $\eta$ and compute the variance to isolate the explainable variance of $\eta_\mathrm{obs}$, finding:
\begin{align}
\Var_\mathrm{exp}(\eta_\mathrm{obs}) ={}
    & \Var(\delta_{\log_{10}F_\mathrm{H\alpha}}) \nonumber\\
    & + \Var(\delta_{\log_{10} F_\mathrm{NUV}}) \nonumber\\
    & + [0.4(k_\mathrm{H\alpha}Q_{sg}^{-1} - k_\mathrm{NUV})]^2 \Var(\delta_{\mathrm{E(B-V)_{stars}}}) \nonumber\\
    & + [0.4 k_\mathrm{H\alpha}]^2\Var(\mathrm{E(B-V)_{stars}}\delta_{Q_{sg}^{-1}}) \nonumber \\
    & + [0.4k_\mathrm{H\alpha}]^2\Var(\delta_\mathrm{Q_{sg}^{-1}}\delta_\mathrm{E(B-V)_{stars}}) \nonumber\\
    & - 2 \Cov(\delta_{\log_{10}F_\mathrm{H\alpha}},\delta_{\log_{10} F_\mathrm{NUV}}) \nonumber\\
    & + 2 [0.4(k_\mathrm{H\alpha}Q_{sg}^{-1} - k_\mathrm{NUV})]\Cov(\delta_{\log_{10}F_\mathrm{H\alpha}}, \delta_{\mathrm{E(B-V)_{stars}}}) \nonumber\\
    & + 2 [0.4k_\mathrm{H\alpha}]\Cov(\delta_{\log_{10}F_\mathrm{H\alpha}}, \mathrm{E(B-V)_{stars}}\delta_{Q_{sg}^{-1}}) \nonumber \\
    & + 2(0.4k_\mathrm{H\alpha}) \Cov(\delta_{\log_{10}F_\mathrm{H\alpha}},\delta_\mathrm{Q_{sg}^{-1}}\delta_\mathrm{E(B-V)_{stars}}) \nonumber\\
    & - 2 [0.4(k_\mathrm{H\alpha}Q_{sg}^{-1} - k_\mathrm{NUV})]\Cov(\delta_{\log_{10} F_\mathrm{NUV}}, \delta_\mathrm{E(B-V)_{stars}}) \nonumber \\
    & - 2 [0.4k_\mathrm{H\alpha}]\Cov(\delta_{\log_{10}F_\mathrm{NUV}}, \mathrm{E(B-V)_{stars}}\delta_{Q_{sg}^{-1}}) \nonumber \\
    & - 2[0.4k_\mathrm{H\alpha}] \Cov(\delta_{\log_{10}F_\mathrm{\nu, NUV}},\delta_\mathrm{Q_{sg}^{-1}}\delta_\mathrm{E(B-V)_{stars}}) \nonumber\\
    & + 2 [0.16k_\mathrm{H\alpha}(k_\mathrm{H\alpha}Q_{sg}^{-1} - k_\mathrm{NUV})] \Cov(\delta_\mathrm{E(B-V)_{stars}}, \mathrm{E(B-V)_{stars}}\delta_{Q_{sg}^{-1}}) \nonumber \\
    & + 2 [0.16k_\mathrm{H\alpha}(k_\mathrm{H\alpha}Q_{sg}^{-1} - k_\mathrm{NUV})] \Cov(\delta_\mathrm{E(B-V)_{stars}}, \delta_\mathrm{E(B-V)_{stars}}\delta_{Q_{sg}^{-1}}) \nonumber \\
    & + 2 [0.4k_\mathrm{H\alpha}]^2 \Cov(\mathrm{E(B-V)_{stars}}\delta_{Q_{sg}^{-1}}, \delta_\mathrm{E(B-V)_{stars}}\delta_{Q_{sg}^{-1}}).
\end{align}
Again, each of the covariance terms without a repeated  operand (i.e., $\Cov(\delta_X,\delta_Y$ or $\Cov(\delta_X,\delta_Y\delta_Z)$) can be set to zero for similar reasons to the previous case.  The fifth variance term can be expanded using the relationship $\Var(\delta_X\delta_Y) = \E(\delta_X)^2\Var(\delta_Y) + \E(\delta_Y)^2\Var(\delta_X) + \Var(\delta_X)\Var(\delta_Y)$ under the assumption that $\delta_X$ and $\delta_Y$ are independent, and all covariance terms with repeated operands will reduce to $\Cov(\delta_X,\delta_X\delta_Y)=\Var(\delta_X)\E[\delta_Y]$ and $\Cov(\delta_X\delta_Y, \delta_X\delta_Z)=\Var(X)\E[\delta_Y]\E[\delta_Z]$ under a similar independence assumption between all terms.  This assumption is possible because none of the measured values (e.g., $\mathrm{E(B-V)}$) are strongly dependent on any single observable (particularly when measured using SED fitting).  This gives:
\begin{align}
\Var_\mathrm{exp}(\eta_\mathrm{obs}) ={}
    & \Var(\delta_{\log_{10}F_\mathrm{H\alpha}}) \nonumber\\
    & + \Var(\delta_{\log_{10} F_\mathrm{NUV}}) \nonumber\\
    & + [0.4(k_\mathrm{H\alpha}Q_{sg}^{-1} - k_\mathrm{NUV})]^2 \Var(\delta_{\mathrm{E(B-V)_{stars}}}) \nonumber\\
    & + [0.4 k_\mathrm{H\alpha}]^2\Var(\mathrm{E(B-V)_{stars}}\delta_{Q_{sg}^{-1}}) \nonumber \\
    & + [0.4k_\mathrm{H\alpha}]^2 \E[\delta_\mathrm{E(B-V)_{stars}}]^2\Var(\delta_{Q_{sg}^{-1}}) \nonumber \\
    & + [0.4k_\mathrm{H\alpha}]^2 \E[\delta_{Q_{sg}^{-1}}]^2\Var(\delta_\mathrm{E(B-V)_{stars}}) \nonumber \\
    & + [0.4k_\mathrm{H\alpha}]^2 \Var(\delta_{Q_{sg}^{-1}})\Var(\delta_\mathrm{E(B-V)_{stars}}) \nonumber \\
    & + 2 [0.16k_\mathrm{H\alpha}(k_\mathrm{H\alpha}Q_{sg}^{-1}-k_\mathrm{NUV})]\E[\delta_{Q_{sg}^{-1}}]\Var(\delta_\mathrm{E(B-V)_{stars}}) \nonumber \\
    & + 2[0.4k_\mathrm{H\alpha}]^2\E[\mathrm{E(B-V)_{stars}}]\E[\delta_\mathrm{E(B-V)_{stars}}]\Var(\delta_{Q_{sg}^{-1}})
\end{align}

Because the distribution of observational errors in these measured quantities is assumed to be approximately symmetric, all expectation values of $\delta$ terms above can be set to zero, and we can reformulate the equation in terms of the reported uncertainties to get an analogous expression to Equation \ref{Eqn: scatter_balmer} for the case of unknown Balmer decrements:
\begin{align}
\Var_\mathrm{exp}(\eta_\mathrm{obs}) ={}
    & \Avg(\sigma^2_{\log_{10}F_\mathrm{H\alpha}}) \nonumber\\
    & + \Avg(\sigma^2_{\log_{10} F_\mathrm{NUV}}) \nonumber\\
    & + [0.4(k_\mathrm{H\alpha}Q_{sg}^{-1} - k_\mathrm{NUV})]^2 \Avg(\sigma^2_{\mathrm{E(B-V)_{stars}}}) \nonumber\\
    & + [0.4 k_\mathrm{H\alpha}]^2\Avg(\mathrm{E(B-V)_{stars}}^2\sigma^2_{Q_{sg}^{-1}}) \nonumber \\
    & + [0.4k_\mathrm{H\alpha}]^2 \Avg(\sigma^2_{Q_{sg}^{-1}})\Avg(\sigma^2_\mathrm{E(B-V)_{stars}}).\label{Eqn: scatter_no_balmer}
\end{align}
Here, $\sigma_{Q_{sg}^{-1}}^2$ represents the variance of $Q_{sg}^{-1}$ as measured from the MOSDEF Gold sample, which we use to approximate the additional measurement uncertainty introduced by the assumption that all galaxies have $Q_{sg} = 0.43$.

In summary, Equation \ref{Eqn: scatter_balmer} describes the relationship between the explainable scatter in $\eta_\mathrm{obs}$ and observed uncertainties in the case of independent measurements of nebular and stellar dust attenuation.  Equation \ref{Eqn: scatter_no_balmer}, on the other hand, describes the relationship for a sample with no independent measurements of stellar and nebular dust, but a known $Q_{sg}$ distribution.  Correspondingly, we will apply Equation \ref{Eqn: scatter_no_balmer} to all three samples using the $Q_{sg}$ distribution measured from the MOSDEF Gold sample.

\begin{deluxetable*}{c|ccc|c}
\tablecaption{Variance Subtraction Results}
\tablehead{\colhead{Sample} & \colhead{$\Var_\mathrm{obs}(\eta_\mathrm{obs})$} & \colhead{$\Var_\mathrm{exp}(\eta_\mathrm{obs})$} & \colhead{$\Var_\mathrm{int}(\eta)$} & 
\colhead{$\boldsymbol{\sigma}_\mathbf{int}(\boldsymbol{\eta})$}}
\startdata
3D-HST & 0.043 & 0.034 & 0.010 & \textbf{0.100} \\
FMOS-COSMOS & 0.058 & 0.054 & 0.004 & \textbf{0.063} \\
MOSDEF & 0.073 & 0.047 & 0.026 & \textbf{0.161}\\
\enddata
\tablecomments{\label{Tbl: eta_var} A table of the observed variance in the $\eta_\mathrm{obs}$, explained variance in $\eta_\mathrm{obs}$ due to uncertainties and scatter in $Q_{sg}$, and the corrected intrinsic variance in $\eta$ after subtracting away the explained variance.}
\end{deluxetable*}

\section{Results} \label{Sec: Results}

\subsection{Mock Catalog Approach}

With the completion of the mock catalog, we can now compare the observed distribution of the burst indicator $\eta$ against that of the mock catalogs, which we show in Figure \ref{fig:all_burstiness_hists}.  Here, we use the Normalized Median Absolute Deviation (NMAD) to estimate the $\eta$ scatter in each panel because it is robust to outliers.  We find that small, but significant differences can be seen when comparing the observed histograms against those of the mock catalogs for each sample.  We find an over-estimation of the observed burstiness for the 3D-HST sample, but an under-estimation of the burstiness for the FMOS-COSMOS and MOSDEF samples.  Further, while the median of $\eta$ is in good agreement with observations for the 3D-HST mock catalog, FMOS-COSMOS and MOSDEF show an under-estimation in the mock catalogs.  We summarize the statistics of each $\eta$ distribution in Table \ref{Tbl: eta_stats}.

\subsection{Variance Estimation Approach}

We apply Equation \ref{Eqn: scatter_no_balmer} to each of our three observed samples while assuming a constant value of $Q_{sg}=0.43$.  We expect that each of these samples should produce differing initial estimates for the total scatter in $\eta$ while producing similar estimates of the intrinsic scatter.  This process yields intrinsic burstiness measurements of 0.100 dex for 3D-HST, 0.063 dex for FMOS-COSMOS, and 0.161 dex for MOSDEF.  A full accounting of the observed scatter, explained scatter, and corrected intrinsic scatter in $\eta$ is given in Table \ref{Tbl: eta_var}.


\section{Discussion} \label{Sec: Discussion}

\subsection{Mock Catalog Offsets}

We find good agreement between the shape of observed and mock $\eta$ histograms for the 3D-HST sample; however, we note a discrepancy between distribution medians for MOSDEF and FMOS-COSMOS observations and mock samples (see Table \ref{Tbl: eta_stats} for detailed statistics).  While this discrepancy could be relieved by assuming $Q_{sg} \approx 1$ (see \citealt{Broussard2019} for illustrations), this would contradict the measurements of $Q_{sg}$ carried out in Section \ref{Sec: Qsg}.  One possible explanation of this behavior could be that high-SNR MOSDEF galaxies do not have $Q_{sg}$ values that are representative of more typical galaxies in MOSDEF and FMOS-COSMOS.  It is also possible that increased precision is needed when measuring $\mathrm{E(B-V)_{stars}}$ and $\mathrm{E(B-V)_{gas}}$ to characterize the $Q_{sg}$ distribution.

\subsection{Ionizing Photon Production Efficiency}

Another potential explanation for this phenomenon could arise from the ionizing photon production efficiency ($\xi$).  Also known as the Lyman Continuum (LyC) production efficiency represented by $\xi_\mathrm{ion}$, the ionizing photon production efficiency is defined as the ratio of the production rate of ionizing photons (denoted as $N(\mathrm{H^0})$) to the UV continuum luminosity density $L_{\nu,\mathrm{UV}}$:
\begin{align}
    \xi_\mathrm{ion} = \frac{N(\mathrm{H^0})}{L_{\nu,\mathrm{NUV}}}.
\end{align}
\cite{Shivaei2018} calculates $N(\mathrm{H^0})$ by assuming Case B recombination, effectively converting the entire H$\alpha$ luminosity into ionizing photons.  They use the relation of \cite{Leitherer1995} to perform this conversion:
\begin{align}
    N(\mathrm{H^0}) = \frac{10^{12} L_\mathrm{H\alpha}}{1.36}
\end{align}
This indicates that $\xi_\mathrm{ion}$ can in fact be expressed as a ratio of the H$\alpha$ luminosity to the UV continuum luminosity density multiplied by a constant.  Thus, the relationship between $\xi_\mathrm{ion}$ and $\eta$ is simply:
\begin{align}
    \eta = {}& \log_{10}\left( \xi_\mathrm{ion} \right) + \log_{10}(C)\\
    C = {}& \frac{C_\mathrm{H\alpha}}{C_{NUV}} \times \frac{1.36}{10^{12}}
\end{align}
As a result, measurements of $\eta$ are subject to the same systematic effects that should be accounted for when measuring $\xi_\mathrm{ion}$.  \cite{Shivaei2018} note four primary sources of uncertainty in measuring $\xi_\mathrm{ion}$, including dust attenuation within HII regions, the escape fraction of LyC photons, and the dust corrections applied to the observed H$\alpha$ luminosities and UV luminosity densities respectively.  Dust attenuation within HII regions has the potential to confound both analyses, but is assumed to be a negligible effect because of the strong agreement between H$\alpha$ and UV SFRs.  Although the LyC escape fraction is expected to be $f_\mathrm{exc}<<1$, a non-zero LyC escape fraction could also bias both analyses, and is unaccounted for in this work.  Finally, the dust attenuation of H$\alpha$ and UV luminosities is accounted in a similar manner by \cite{Shivaei2018} who use the Balmer decrement to correct for nebular dust attenuation and SED fitting to correct for UV dust attenuation.  The ionizing photon efficiency is also dependent on recent star formation and, while an analysis measuring $\xi_\mathrm{ion}$ would typically average over these variations as a systematic effect, we aim to measure them because of their close relationship to burstiness.

\subsection{Variance Estimation Approach}

When we study the contributions of each of the terms in Equation \ref{Eqn: scatter_no_balmer} in practice, the fourth term dominates the correction factor across all three samples, causing the magnitude of the correction factor to strongly depend upon each sample's $\mathrm{E(B-V)}$ values. This causes the FMOS-COSMOS sample's correction factor to rise above those of the other two samples, as it contains galaxies that are, on average, dustier than the others.

While Table \ref{Tbl: eta_var} shows a wide range in measurements of the intrinsic variance in $\eta$, this is largely due to large uncertainties in measurements of Balmer decrements and $\mathrm{E(B-v)_{stars}}$ (and therefore also $Q_{sg}$).  We estimate the range of the intrinsic burstiness to be $0.06-0.16$ dex based on the results of our three observed samples.  Similar to our discussion of the mock catalog results, these measurements will likely benefit from a larger sample of galaxies as well as increased precision for measurements of the $Q_{sg}$ distribution.  Early \textit{JWST} surveys will greatly benefit similar analyses in the future, as the Cosmic Evolution Early Release Science (CEERS) program alone, for example, promises deep spectra of several hundred galaxies at 5$\sigma$ emission line limiting fluxes of $1-2\times10^{-18}~\mathrm{erg~s^{-1}~cm^{-2}}$.


\section{Conclusions}\label{Sec: Conclusions}

This work adds spectroscopy from FMOS-COSMOS and MOSDEF to the 3D-HST dataset analyzed in \cite{Broussard2019}.  Most importantly, the addition of MOSDEF carries with it galaxies with simultaneous H$\alpha$ and H$\beta$ detections, enabling Balmer decrement dust corrections of nebular emission lines.  Combined with updated SED fits using the Dense Basis \citep{Iyer2019} method, we are able to measure the distribution of $Q_{sg}$, finding an average value of $Q_{sg} = 0.43$ and $\sigma_{Q_{sg}} = 0.33$.

We generate updated mock galaxy catalogs using the Santa Cruz SAM and {\sc Mufasa} cosmological hydrodynamical simulations based on the individual observational constraints for the 3D-HST and FMOS-COSMOS surveys, as well as the MOSDEF sample (those galaxies with a SFR$_\mathrm{H\beta}$ SNR$>5$ for which we assume $Q_{sg} = 0.43$, similar to the other two samples).  We find decent agreement between the $\eta$ distributions of the observed samples and those of their respective mock catalogs; however, the FMOS-COSMOS and MOSDEF mocks exhibit a shift toward lower $\eta$ and decreased scatter in $\eta$.  These differences could have a number of causes, including differences in the ionizing photon production efficiency $\xi_\mathrm{ion}$, insufficiently precise measurements of the distribution of $Q_{sg}$, and of course genuine differences between simulations and observations that have not previously been probed.  $Q_{sg}$ measurement precision could be improved by additional spectroscopy targeting Balmer decrements as well as more precise measurements of the stellar dust attenuation.

We also describe the close relationship between $\eta$ and the logarithm of the ionizing photon efficiency $\xi_\mathrm{ion}$, which are related by a constant.  Most of the systematics that must be accounted for in an analysis studying $\xi_\mathrm{ion}$ are also present when measuring $\eta$; meanwhile, we note that variations in recent star formation that are a systematic for $\xi_\mathrm{ion}$ measurements are the target of study of this work.

Beyond the mock catalog approach first introduced in \cite{Broussard2019}, this work adds a novel analysis in which the intrinsic variance in $\eta$ can be estimated by accounting for the amount of scatter in $\eta_\mathrm{obs}$ that is the result of observational uncertainties and variations in $Q_{sg}$.  We are able to perform the first measurement of intrinsic scatter in $\eta$, finding an intrinsic scatter of $0.06-0.16$ dex.  We conclude that more precise measurements of $Q_{sg}$ via stellar dust attenuation or Balmer decrements are likely necessary to improve the precision of this result.  We note early \textit{JWST} surveys will provide more numerous and more precise measurement uncertainties for H$\alpha$ and H$\beta$ that will greatly benefit similar analyses in the near future.

\section*{Acknowledgements}
AB and EG acknowledge support from NASA ADAP grant 80NSSC22K0487, HST grant HST-GO-15647.020-A, and the U.S. Department of Energy, Office of Science, Office of High Energy Physics Cosmic Frontier Research program under Award Number DE-SC0010008.  The authors would also like to acknowledge guidance from Alice Shapley and Bahram Mobasher for guidance on the MOSDEF public data release.

\bibliographystyle{apj_url}
\bibliography{bibliography}

\end{document}